\documentclass[leqno]{article}

\usepackage{amssymb}
\usepackage{amsmath}

\usepackage[margin=2cm]{geometry}
\usepackage{parskip}

\begin{document}

\title{Estimating Bernoulli trial probability from a small sample}

\author{Norman D.~Megill\\ Boston Information Group, 19 Locke Ln., Lexington,
MA 02420, USA\\ \\and\\ \\ Mladen Pavi\v ci\'c\\ Chair of Physics, Faculty of Civil Engineering,
University of Zagreb, Croatia}

\bibliographystyle{unsrt}
\pagenumbering{arabic}
\pagestyle{myheadings}

\maketitle

\abstract{
The standard textbook method for estimating the probability of a biased
coin from finite tosses implicitly assumes the sample sizes are large
and gives incorrect results for small samples.  We describe the exact
solution, which is correct for any sample size.}


\section{Introduction}\label{sec:intro}

Consider the following problem.  A biased coin, with an unknown
probability $p$ of heads, is tossed $n$ times, and $m$ heads result.
What is the best estimate of $p$ from $n$ and $m$?

Problems of this form occur in many applications.  A typical example is
found in Ref.~\cite[p.~346]{statistics}.  Their problem is to determine
the percentage of Democratic votes, and the confidence interval for that
percentage, given that 917 voters in a sample of 1,600 (out of 25,000)
are Democrats.

The solution given in the book is as follows.  The observed ratio
$\frac{917}{1,600}\approx 0.57$ is the estimate for the fraction of
Democratic voters.  The standard deviation (SD) is estimated by a
bootstrap procedure as $\sqrt{0.57\cdot 0.43} \approx 0.50$.  The standard
error (SE) is computed as $\sqrt{1,600}\cdot \mathrm{SD} \approx 20$, or
$1.25\%$.  The $95\%$ confidence interval is estimated as $57\pm 2\cdot
1.25\%$.

This method, which we will call the ``standard method,'' does not work
when samples sizes are small or the fraction is near the extremes of 0
or 1. For example, suppose an urn is filled with marbles, an unknown
fraction of which are red and the rest white.  A sample of 5 marbles is
taken (with replacement), and in that sample all marbles are white.
What is the fraction and $80\%$ confidence interval of red marbles?  The
standard method yields a fraction of zero and a confidence interval of
zero.  These are obviously wrong.  A sample of 5 whites indicates that
the fraction of reds is probably small, but it certainly provides no
assurance that it is zero; instead, it is very likely to be nonzero.

We will describe another method, which we will call the ``exact
method,'' that does not have these errors.  There are many practical
cases where small sample sizes are important.  For example, a medical
trial may involve just a dozen or so patients.  It can be useful to use
the exact method for such studies.  Indeed, in some cases it might be
considered irresponsible not to use it, since the standard method could
lead to incorrect decisions based on misleading results.

This problem arose when the authors needed an estimate and confidence
interval for the probability of a rare event that may occur less than a
dozen times, or even never, in a sample of several billion.~\cite{mfwap-11}
After the standard method yielded obviously wrong
results, the authors were surprised that a literature search did not
yield the exact method.  The purpose of this note is to document
it for general use.

The next section describes the assumptions and detailed derivation of
the exact method.  The reader who just wants to see the final result
may refer to Eqs.~(\ref{expected}), (\ref{conf1}), and (\ref{conf2})
below, which show the exact method's mean, lower confidence level,
and upper confidence level respectively.

\section{The exact method}

The primary example we will use in our development is the biased coin
problem, which is equivalent to a finite sample from an infinite
population.  Formally, the problem is to estimate an unknown probability
of success $p$ in a Bernoulli process, knowing only that there were $m$
successes in an experimental run of $n$ trials.

The equivalent problem for finite populations is sampling with
replacement, where each sample is put back into the the population pool
so that it will have an equal chance of being drawn again.

Let $p$ be the unknown probability of heads for the biased coin.  We
will assume that $p$ is uniformly distributed between 0 and 1, that is,
all values of $p$ between 0 and 1 are equally likely.  This seems to be
a reasonable assumption absent any other information.

We will first look at the case where $p$ has $k$ discrete
values between 0 and 1. This will let us study the problem with
simple examples in order to understand the sample space intuitively.
(The discrete case can also stand on its own as a useful result when
whenever the probabilities actually are discrete.)
Once we derive the result for arbitrary $k$, we can take the limit
as $k\to \infty$ to obtain the exact result for a continuously distributed
$p$.

We will call the
discrete values of $p$ by $p_1, p_2,\ldots,p_k$:
\begin{align}
p_1 &= \frac{1}{k}(1 - \frac{1}{2})
           &&\text{(representing $0 \le p < \frac{1}{k}$)}  \notag \\
p_2 &=  \frac{1}{k}(2 - \frac{1}{2})
        &&\text{(representing $\frac{1}{k}k \le p < \frac{2}{k}$)}  \notag \\
     & \cdots                                           \notag \\
p_k &=  \frac{1}{k}(k - \frac{1}{2})
          &&\text{(representing $\frac{k-1}{k} \le p \le 1$)}.  \notag
\end{align}

Given a coin with probability $p_i$ ($1\le i\le k$) of heads, the
probability of tails is $1-p_i$.  The probability of a specific finite
sequence beginning (for example) head, tail, head, head tail,\ldots is
$p_i(1-p_i)p_i p_i(1-p_i)\ldots$.  The probability of obtaining a {\em
specific} sequence of $n$ tosses containing $m$ heads is thus
\begin{align}
  p_i^m  (1-p_i)^{n-m}.       \notag
\end{align}

There are $\binom{n}{m} = \frac{n!}{m!(n-m)!}$ ways of obtaining $m$
heads out of $n$ tosses.  Thus the probability of exactly $m$ heads in
$n$ tosses is
\begin{align}
  \binom{n}{m}  p_i^m  (1-p_i)^{n-m}.   \notag
\end{align}

To motivate the main argument, consider the simple example where $k=2$.
We have:
\begin{align}
  p_1 &= \frac{1}{4},  &p_2 = \frac{3}{4}.     \notag
\end{align}
Suppose we perform a large number $t$ of trials (which we can later
take to infinity---actually, $t$ will cancel in the final
result), say $t=1,000,000$, each with $n$ tosses, for a coin
with probability $p_1$ and also for a coin with probability $p_2$.  The
expected number of $n$-toss trials resulting in $m$ heads will be
\begin{align}
   q_1 + q_2     \notag
\end{align}
where
\begin{align}
   q_1 &= t  \binom{n}{m}  p_1^m  (1-p_1)^{n-m} \notag \\
   q_2 &= t  \binom{n}{m}  p_2^m  (1-p_2)^{n-m}.  \notag
\end{align}
Thus for any particular $n$-toss trial with $m$ heads, the probability
that it came from the $p_1$ coin is
\begin{align}
  e_1 &= \frac{q_1}{q_1+q_2} =
            \frac{p_1^m (1-p_1)^{n-m}}{p_1^m  (1-p_1)^{n-m}
                +  p_2^m  (1-p_2)^{n-m}}    \notag
\end{align}
and similarly for the $p_2$ coin.  As an example, for n=5 and m=1,
we have
\begin{align}
  e_1 &\approx 0.964,  & e_2 \approx 0.036.     \notag
\end{align}
This means that if we know that a coin has an
unknown probability of heads of
either $\frac{1}{4}$ or $\frac{3}{4}$, and we observe 1 head in a 5-toss
sample, $96.4\%$ of the time the coin's probability will be
$\frac{1}{4}$.

Going back to the general case, the expected probability of a coin with
head probability $p_i$, $1 \le i \le k$, based on a sample of $n$ tosses
where $m$ heads are observed, is
\begin{align}
  e_i &= \frac{p_i^m (1-p_i)^{n-m}}{\sum_{j=1}^k {p_j^m  (1-p_j)^{n-m}}}.
             \notag
\end{align}
The mean expected probability is computed in the standard way:
\begin{align}
  E[p_i] = \sum_{i=1}^k {\frac{p_i}{k} e_i}    \notag
\end{align}
and the confidence interval can be computed (say with a computer
algorithm) from the distribution $e_i$.  The number of intervals $k$ can
be made as large as desired for sufficient accuracy.

We take the limit as $k\to\infty$ to obtain the exact (continuous)
probability density $e(x)$ for head probability $x$, $0\le
x\le 1$:
\begin{align}
e(x) &=\frac{x^m  (1-x)^{n-m}}{\int_0^1 {y^m  (1-y)^{n-m} dy}}
   = \frac{x^m  (1-x)^{n-m}}{B(m+1, n-m+1)}     \notag
\end{align}
where $B(i,j)\equiv \int_0^1{y^{i-1}(1-y)^{j-1} dy}$ is the beta function.

The exact expectation value of the distribution $e(x)$ is then
\begin{align}
  E[e(x)] &= \frac{\int_0^1 x  [x^m  (1-x)^{n-m}] dx}{B(m+1, n-m+1)} =
      \frac{B(m+2, n-m+1)}{B(m+1, n-m+1)}. \notag
\end{align}
It can be shown that the last ratio evaluates to $\frac{m+1}{n+2}$.
Thus we have a surprisingly simple formula for the expected probability
of heads for the biased coin,
\begin{align}
  E[e(x)] &= \frac{m+1}{n+2} \label{expected}.
\end{align}
(This compares to the corresponding standard method
expectation $\frac{m}{n}$,
showing the two are nearly the same for large $n$ and $m$.)

The confidence interval is a little harder to compute.
For a confidence interval of $c\cdot 100\%$, we need to find
$x_1$ and $x_2$ such that the cumulative distribution
of the probability density $e(x)$
equals $\frac{1}{2}(1-c)$ and $\frac{1}{2}(1+c)$, for example
$0.1$ and $0.9$ for an $80\%$ confidence interval.
\begin{align}
 \int_0^{x_1}e(x)dx &= \frac{\int_0^{x_1}x^m  (1-x)^{n-m}dy}{B(m+1, n-m+1)}
    =\frac{1}{2}(1-c)        \notag
\\
 \int_0^{x_2}e(x)dx &= \frac{\int_0^{x_1}x^m  (1-x)^{n-m}dy}{B(m+1, n-m+1)}
    =\frac{1}{2}(1+c).     \notag
\end{align}
The integrals can be expressed with regularized incomplete beta functions
$I_{x_1}(m+1,n-m+1)$ and $I_{x_2}(m+1,n-m+1)$, so obtaining the
confidence interval amounts to solving the two equations
\begin{align}
I_{x_1}(m+1,n-m+1)  &=  \frac{1}{2}(1-c)    \notag \\
I_{x_2}(m+1,n-m+1)  &=  \frac{1}{2}(1+c))  \notag
\end{align}
for $x_1$ and $x_2$.
The solutions can be expressed as inverse regularized
incomplete beta functions:
\begin{align}
x_1=I^{-1}_{\frac{1}{2}(1-c)}(m+1,n-m+1)
       \label{conf1} \\
x_2=I^{-1}_{\frac{1}{2}(1+c)}(m+1,n-m+1)
       \label{conf2}
\end{align}
These can be evaluated using, for example, a computer algebra
system.\footnote{$I_{1-p}(n-k,k+1)$ is
the well-known cumulative distribution function for the number of
successes $k$ for the binomial distribution of $n$ trials from a
Bernoulli process with a {\em known} probability $p$ of success.
Because this problem has frequent applications,
most computer algebra systems provide the regularized
incomplete beta function and its inverse.}

{\em Example 1.} For the marble problem described in
Sec.~\ref{sec:intro}, we have $c=0.8$, $m=0$, $n=5$.  The exact
method shows that the $80\%$ confidence interval is between
$x_1\approx 0.017$ and $x_2\approx 0.319$, with a mean from
Eq.~\ref{expected} of $\frac{1}{7}\approx 0.143$.  This is very
different from the (incorrect) zero confidence interval and zero mean
that the standard method yields.

{\em Example 2.} For the voting problem described in
Sec.~\ref{sec:intro}, we have $c=0.95$, $m=917$, $n=1600$.  With the
exact method, the mean is $\frac{918}{1602}\approx 0.573$ with a
$95\%$ confidence interval between $x_1\approx 0.549$ and $x_2\approx
0.597$.  The standard method yields mean $0.573$ and $95\%$ confidence
interval between $0.548$ and $0.598$, showing that the two methods
approximately agree when the sample size is large.


\section{Conclusion}

The formulas for the exact method are nearly as simple to state as those
for the standard method.  But they have the significant advantage of
being exact rather than approximate, with no errors when sample sizes
are small.

In the exact method, the derivation from first principles is
straightforward and rigorous, with all assumptions clearly laid out.
This contrasts to the standard method, which involves the mathematically
questionable (or at least not rigorously justified) bootstrapping
procedure as well as the implicit use of Gaussian distributions to
approximate non-Gaussian ones.  The errors involved in these
approximations, as well as their their regimes of validity, are
difficult to determine and typically glossed over.  Regarding
bootstrapping, the authors say merely that ``the estimate is good when
the sample is reasonably large'' even though the procedure ``may seem
crude.''  \cite[p.~342]{statistics}

It is not clear why the exact method isn't mentioned in most textbooks
or, indeed, why it isn't universally used instead of the standard
method.  Apparently the exact method is not well known.

\end{document}